# Nanoscale ordered Layered-Segregation of Cobalt and Iron in the Perovskite SmFe$_{0.5}$Co$_{0.5}$O$_3$

Nathalia E.Mordvinova[1], Ashish Shukla[2], Asish K. Kundu[2], Oleg I. Lebedev[1], Md. Motin Seikh[1,3], Vincent Caignaert[1] and Bernard Raveau[1]

[1]*Laboratoire CRISMAT ENSICAEN-CNRS, University of Normandy 14000 Caen, France*
[2]*Discipline of Physics, Indian Institute of Information Technology, Design & Manufacturing Jabalpur, Dumna Airport Road, Madhya Pradesh–482005, India*
[3]*Department of Chemistry, Visva-Bharati University, Santiniketan, West Bengal –731235, India*

Iron and cobalt perovskites form a huge class of materials, which have been investigated for their attractive magnetic, transport and oxygen storage properties making them potential materials for applications, as exemplified by the layered perovskites LnBaCo$_2$O$_{5+\delta}$,[1-25] LnBaFe$_2$O$_{5+\delta}$.[26–32] Considering the ability of these two cations to occupy the same B sites of the ABO$_{3-\delta}$ perovskite structure, the synthesis of mixed "Fe-Co" perovskites offers a vast field for the realization of new properties, which has been incompletely explored. The possibility to induce ordering between two different transition elements on the B sites of a perovskite structure ABO$_{3-\delta}$ is an important issue which has not been addressed clearly to date. This is illustrated by the recent observation of an exceptional layered ordering between cobalt and iron in the oxygen deficient quintuple perovskite Y$_2$Ba$_3$Fe$_3$Co$_2$O$_{13+\delta}$($\delta\sim$0.36).[33] The latter exhibits a unique nanostructure involving a double layered cationic ordering of both Y/Ba and Fe/Co layers and simultaneously of oxygen vacancies corresponding to the ideal intergrowth [YBaCo$_2$O$_5$][YBa$_2$Fe$_3$O$_8$]. Thus, its structure can be considered as a consequence of the stability of two classes of layered perovskites: the double "112" perovskites LnBaCo$_2$O$_{5+\delta}$,[1–25] and LnBaFe$_2$O$_{5+\delta}$,[26–32] synthesized for both cobalt and iron and the triple "123" perovskites LnBa$_2$Fe$_3$O$_{8+\delta}$,[34–38] obtained only for iron. This behavior is different from that of other quintuple perovskites Ln$_2$Ba$_3$Fe$_{5-x}$Co$_x$O$_{15-\delta}$ with Ln = Sm, Eu, Nd,[39-41] which do not show any layered ordering of iron and cobalt in spite of the presence of ordered Ln/Ba layers.

These results suggest that the Co$^{3+}$ and Fe$^{3+}$ cations, in spite of their similar ability to accommodate both the octahedral and pyramidal coordination, exhibit a different distortion of their oxygen polyhedra, which favors a possible ordering. Such a layered separation of cobalt and iron is induced in these phases by the nature and ordering of the Ln/Ba cations and by the oxygen stoichiometry and ordering. This raises the issue of the stability of a possible intrinsic ordering of cobalt and irons in a stoichiometric perovskite containing only one sort of A cation. Thus, we have tried to generate ordering of Co$^{3+}$ and Fe$^{3+}$ cations in the stoichiometric perovskite SmFe$_{0.5}$Co$_{0.5}$O$_3$ whose two transition metal cations are usually expected to be distributed at random due to their similar size and to their ability to accommodate the octahedral coordination. For this purpose we used the sol gel synthesis process[42] which offers the possibility to control the reaction pathways on a molecular level during the transformation process of the precursors to the final product and to work at lower temperature. In the present study, we compare the nanostructural behavior of two sorts of samples synthesized by direct solid state (DSS) method at high temperature (1100°C) and by sol-gel (SG) technique at lower temperature (950°C) respectively. We show that, when prepared by the sol-gel method, this perovskite exhibits a unique nanostructure built up of mixed Co/Fe octahedral layers involving a gradual and periodic variation of the cationic composition from Co-rich to Fe-rich layers and vice versa. Moreover, the latter grows coherently with the parent matrix, itself characterized by a statistical distribution of the Co$^{3+}$and Fe$^{3+}$ cations, similar to the one obtained by direct synthesis.

The powder X-ray diffraction patterns of both samples, DSS and SG, are absolutely identical (Figure S1 in the Supporting Information), and can be indexed in the same *Pnma* orthorhombic cell, $a_o$=5.484 Å ~ $a_p\sqrt{2}$, $b_o$=7.61 Å~2$a_p$, $c_o$=5.348 Å ~ $a_p\sqrt{2}$, similar to the parent perovskites SmFeO$_3$[43-46] and SmCoO$_3$.[47,48] No superstructure reflections and no secondary phase could be detected for both samples.

The good crystallinity of SmFe$_{0.5}$Co$_{0.5}$O$_3$ DSS and SG samples is attested by HRTEM observations and ED studies. As expected, the [010], [101], [210] and [100] ED patterns of the DSS sample (Figure 1a) can be indexed in the well known orthorhombic *Pnma* structure, in agreement with PXRD results. A large part of the crystallites of the SG samples (called disordered-SG) also exhibit similar ED patterns already observed for the *Pnma* structure. However, a significant number of crystallites (called ordered SG), around 30-40%, show the appearance of superstructure reflections along the $c_o$ ($a_p\sqrt{2}$) direction and indicate a space group different from *Pnma*. For the sake of clarity, the corresponding diffraction patterns of this modulated structure will be labelled [010]$^m$ and [210]$^m$ patterns (Figure 1b). Such additional reflections imply an ordering phenomenon characterized by the *n* x $a_p\sqrt{2}$ periodicity along $c_o$. The modulated character of this ordering centred around the n =6 and 7 members will be further discussed. It must also be emphasized that the ordered and disordered regions often coexist in the same crystallite as illustrated from the comparison of the [210] patterns in Figures 1a and 1b which were obtained for the same crystal.

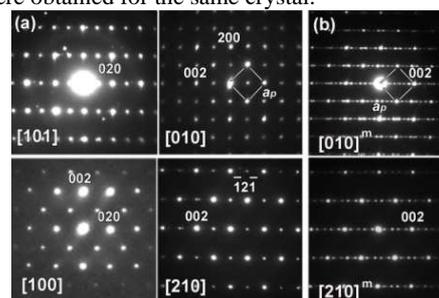

Figure 1 (a) - ED patterns collected along the main crystallographic zones for DSS and "disordered" SG SmFe$_{0.5}$Co$_{0.5}$O$_3$ structures and (b) for "ordered" modulated SG structure. Note the appearance of superstructure spots in the modulated structure along $a_p\sqrt{2}$ direction in [010] and [210] zones.

The low-magnification HAADF-STEM images of the disordered-SG SmFe$_{0.5}$Co$_{0.5}$O$_3$, show that the sample consists of <100> / <001> well faceted nanoparticles with typical mean particle size of about 0.5-2 μm, which are defect-free (Figure 2a). The good crystallinity of this sample and its *Pnma* structure are confirmed by high resolution HAADF-STEM and ABF-STEM imaging along the main zone axis (Figures 2b-d). The HAADF-STEM images along two main zone axis [010] and [101](Figures 2b,2d respectively) show a uniform contrast of atomic columns suggesting an homogeneous distribution of Sm (brighter dots) and Fe/Co (less bright dots) cations within the atomic columns. Importantly, the enlargement of the ABF-STEM image (Figure 2c) does not allow any oxygen vacancy to be detected suggesting that all the anionic sites are fully occupied.

However, the contrast over the single nanocrystallites is not uniform – brighter at the edge and darker at the centre (Figure 2a). Bearing in mind that the contrast at HADDF-STEM is directly proportional to the thickness of the crystal and to the atomic number of the elements (~$Z^2$), the difference in contrast can be explained either by a variation of composition (e.g. Sm rich area) or of thickness (thicker at the edge or less material at the centre of nanoparticles). In order to clarify this point, the energy-dispersive X-ray spectroscopy (EDX) study was performed (Figure 2a colour). The latter shows that the distribution of the Sm, Fe and Co elements is uniform with an atomic ratio close to nominal composition (O/Fe/Co/Sm – 21.17/23.93/21.57/33.32 at %). Thus, it evidences that the thickness is responsible for this effect. This random distribution of Fe and Co was also confirmed by EELS at atomic level, as shown ~~from~~ in the elemental maps for Fe $L_{2,3}$, Co $L_{2,3}$, Sm $M_{4,5}$ and their color overlap along [010] zone axis (Figure 2e) which clearly evidence a uniform distribution of all elements. Only a careful analysis of EELS elemental mapping revealed some columns with higher concentration of Co (darker blue colour in Figure 2e).

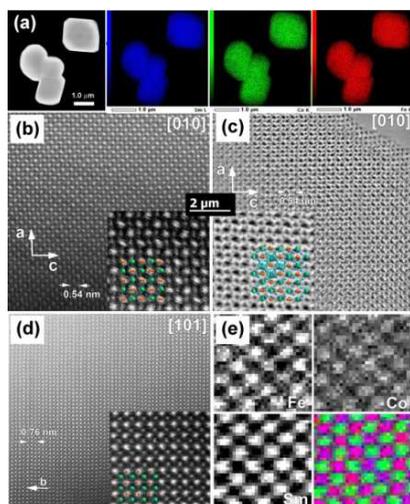

Figure 2 Disordered SG sample: (a) Low magnification HAAD-STEM image and corresponding EDX maps of Sm (blue), Fe (green) and Co (yellow). (b) High resolution [010] HAADF-STEM and (c) corresponding ABF-STEM images; (d) [101] HAADF-STEM Insets show overlaid model (Sm- orange, Fe/Co – green, O – blue); (e) [010]EELS atomic mapping, Fe $L_{2,3}$ map, Co $L_{2,3}$ map, Sm $M_{4,5}$ map (Sm in green, Co in blue and Fe in red).

The TEM studies of the ordered crystals obtained from the sol gel synthesis (ordered-SG) show the coexistence of several superstructures, with possible modulations, as suggested by the preliminary ED investigation (Figure 1b). The $[010]^m$ and $[210]^m$ HAADF-STEM images (Figures 3a-b) show the coexistence of at least two types of ordered layers which are brighter than the Sm layers and are separated from each other by 6 or 7 Sm layers respectively. These periodicities corroborate the corresponding ED patterns (inset FT patterns Figures 3a-b). Such observations explain that the ED patterns obtained for many of these crystals (Figure1a) correspond to the superposition of at least two families of superstructure reflections along the $[110]_P$ or $[001]_{Pnma}$ direction, as reconstructed in Figure 3c. They imply the coexistence of ordered structures characterized by "$a_p\sqrt{2}$" x $2a_p$ x $n$ x $a_p\sqrt{2}$" supercells with n=6 (blue in Figure 3c inset) and n=7 (yellow in Figure 3c inset). Bearing in mind the layered ordering observed for cobalt and iron in the quintuple perovskites $Y_2Ba_3Fe_3Co_2O_{13+\delta}$,[33] the contrasts of Figures3a-b suggest that these cations are also displayed here in the form of ordered layers.

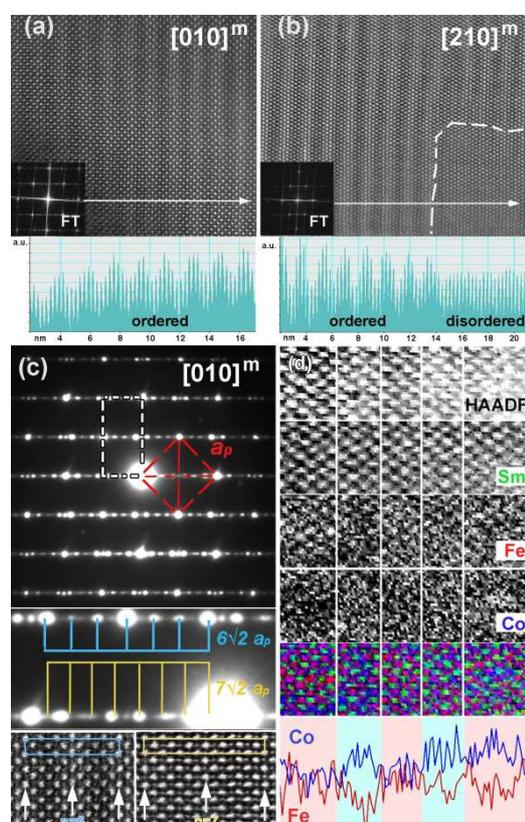

Figure 3. Ordered-SG sample: HR-HAADF-STEM images along $[010]_{Pnma}$ (a) and $[210]_{Pnma}$ (b) zone axis and corresponding intensity plot profile illustrated wavy character of contrast (c) - $[010]^m$. ED pattern showing periodicities corresponding to $6a_p\sqrt{2}$ (blue) and $7a_p\sqrt{2}$ (yellow). Fragments of corresponding HAADF-STEM images are given as inset (bottom panel). (d) EELS mapping of Sm $M_{4,5}$, Fe $L_{2,3}$ and Co $L_{2,3}$ showing segregation and modulation of Co and Fe within the modulated structure. The intensity plot profile for Co (blue) and Fe (red) is given as inset.

However, the corresponding plot profiles (inset Figures 3a-b) show that the ordering of these cations should not consist of pure cobalt and pure iron layers, but rather of mixed Fe/Co layers involving a modulated cationic distribution along the stacking sequence (labelled *c* on Figure 3). The atomic EELS elemental mapping (Figure 3d) allows to definitely conclude about the nature of this cationic ordering. One indeed observes a perfect stacking of the Sm layers, indicating that the latter do not exhibit any compositional variation. Remarkably, the Co content and correlatively the Fe content varies continuously from one layer to the next. This is shown on the coloured micrograph in Figure 3d and corresponding intensity line profile: starting from a cobalt-rich layer (blue colour), mixed Fe/Co layers are formed involving a decrease of the Co content and going finally to a Fe- rich layer (pink colour). The corresponding profile evolution of these cationic contents clearly shows that for a same layer, when the Co content (blue) reaches a maximum, the iron content (pink) is minimum and vice versa. Important to notice that, as for the disordered perovskite, we did not detect any oxygen vacancies from the ABF-STEM studies.

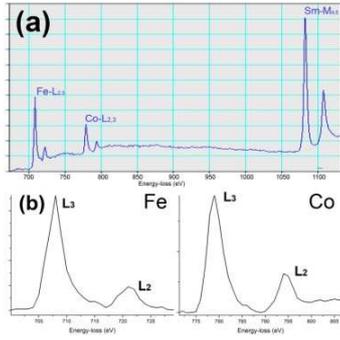

Figure 4: (a) single crystallite EELS fine structure of $SmFe_{0.5}Co_{0.5}O_3$, (b) EELS $L_{2,3}$ edges of Fe and Co.

In order to determine the valence states of iron and cobalt, EELS spectra were registered for both samples, disordered and ordered- SG perovskites. Identical results were obtained for the both compounds. EELS fine structure of these oxides (Figure 4a) shows the Fe-$L_{2,3}$, Co-$L_{2,3}$ and Sm-$M_{4,5}$ edge simultaneously, whereas the EELS of Fe-$L_{2,3}$ and Co-$L_{2,3}$ with energy resolution of ~0.7eV, after background correction are shown in Figure 4b. It is well established that the shape of $L_{2,3}$ edges, the chemical shifts and the $L_3/L_2$ ratio can be used for the determination of the valence state of the transition elements . Detailed analysis of Fe-$L_{2,3}$ and Co $L_{2,3}$ EELS reveal that the $L_3$ positions of these cations are close to 708 eV and 779 eV, respectively, with energy differences ($L_3$-$L_2$) around 13 eV and 15 eV, whereas the $L_3/L_2$ white-line ratios are 4.5 and 2.5 respectively. Comparison with the data reported in the literature for iron[49, 50] and cobalt[51] clearly shows that in our perovskites, iron and cobalt are both in the trivalent state, confirming in this way the absence of oxygen vacancies.

The impact of such a nanoscale layered ordering upon the magnetism of perovskites is an important issue. A comparison of the magnetic behaviour of the DSS samples which cobalt and iron cations are fully disordered with that of the SG samples that contain a substantial amount of ordered layered Co/Fe regions sheds light on this problem. The field cooled (FC) magnetization curves of the two samples, SG-ordered (Figure 5a) and DSS (Figure 5b), though closely related to each other exhibit significant differences. At low temperature (T < 150 K), the M(T) curves (Figure 5) are rather similar for both samples, characterized by a broad maximum at ~70 K, with a sharp decrease of the magnetic moment down to 2 K. Considering the isothermal magnetization M (H) curves within this temperature range (Figure 6), the existence of hysteresis loops at 100 K, without saturation, indicates that both samples may exhibit a canted antiferromagnetic state. Then bearing in mind that at low T, the 4f electron Sm sub-lattice and the 3d electron Fe/Co sub-lattice can be arranged antiparallely,[52] this explains that below 70 K there appears an abrupt decrease of the magnetic moment down to 2 K.

In contrast, at higher temperature (T >150 K), the field cooled (FC) behaviour of the two samples is significantly different. The M(T) curve of the SG sample (Figure 5a)shows a sharp increase of magnetization around 300K with a kink at 210 K, as T decreases, suggesting a transition from a paramagnetic state to a new magnetic

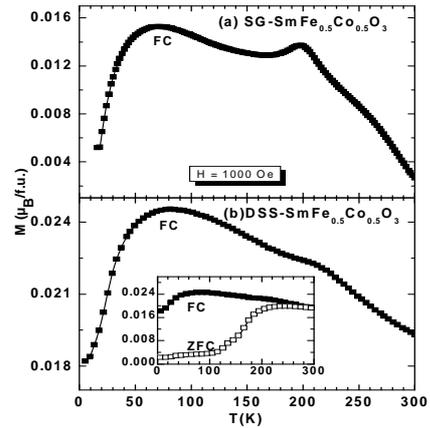

Figure 5: M(T) curves for (a) SG-$SmFe_{0.5}Co_{0.5}O_3$ and (b) DSS-$SmFe_{0.5}Co_{0.5}O_3$ samples ( H=1000 Oe).

state.This hypothesis is corroborated by the corresponding M (H) curves of this sample (Figure 6a) which confirm the paramagnetic nature of the sample at 350 K and show a large hysteresis loop at 100 K suggesting a ferromagnetic like behaviour. Differently, the DSS sample does not show any magnetic transition for T > 150 K (Figure 5b). One only observes a gradual increase of the magnetization as T decreases from 300 K to 100 K. The M (H) curves of this DSS sample (Figure 6b) show narrow hysteresis loops in the whole temperature range even at 350 K. The shape of these loops and the fact that they persist at higher temperature strongly suggest that the nature of the canted antiferromagnetic interactions observed in the DSS sample is different from that of SG sample. Note, however that the M(T) data of the DSS sample (Figure 5b) exhibit a shoulder at 210 K which is reminiscent of the kink observed for the SG sample (Figure5a). Such a feature is most probably related to the existence of some Co/Fe segregation in the DSS sample, as detected from the EELS elemental mapping. It must also be emphasized that the zero field cooled (ZFC) data are different for the two samples: the ZFC curve is superimposed to the FC one for the SG sample, whereas in contrast the DSS sample exhibits a large divergence between the ZFC and FC data (inset Figure 5b).

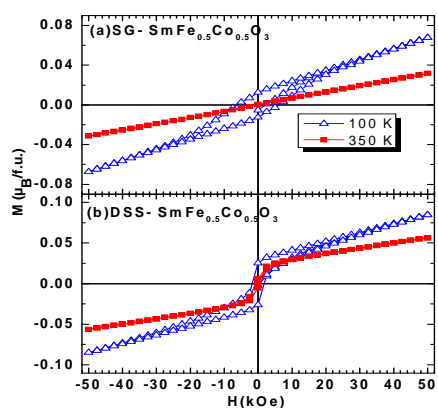

**Figure 6:** M(H) curves at two different temperatures for (a) SG-SmFe$_{0.5}$Co$_{0.5}$O$_3$ and (b) DSS-SmFe$_{0.5}$Co$_{0.5}$O$_3$ samples.

The fact that that the SG and DSS samples exhibit a different magnetic behaviour is closely related to the different distributions of the Fe$^{3+}$ and Co$^{3+}$ cations in the perovskite matrix. In the DSS sample the random distribution of those species, with the 1:1 cationic ratio makes that the antiferromagnetic Fe$^{3+}$-O-Co$^{3+}$ interactions are predominant and consequently control the antiferromagnetism in this disordered perovskite. In contrast, in the SG sample, the concomitant presence of Co$^{3+}$-rich and Fe$^{3+}$-rich layers implies that the substantial presence of Co$^{3+}$-O-Co$^{3+}$ and Fe$^{3+}$-O-Fe$^{3+}$ antiferromagnetic magnetic interactions that will compete with Fe$^{3+}$-O-Co$^{3+}$ interactions. According to literature report by several groups, all these interactions Fe$^{3+}$-O-Co$^{3+}$, Fe$^{3+}$-O-Fe$^{3+}$[40, 42] and Co$^{3+}$-O-Co$^{3+}$[1, 9, 53] are antiferromagnetic in nature. However, due to spin canting the resultant magnetic moment could be different in all these type of interactions.

In conclusion, the possibility to segregate cobalt and iron in a single perovskite, at a nanoscale in the form of layers has been demonstrated, using sol gel technic synthesis. Quite remarkably, it is shown that such a nanoscale ordering, which can only be detected by a combined HAADF-STEM and EELS study, has a significant impact upon the magnetic properties of this oxide. This suggests that careful nanoscale characterization of oxides will be necessary in the future before proceeding to a viable interpretation of their physical properties, at least in the field of magnetism.

**References**
(1) Raveau, B.; Seikh, M. M. Cobalt Oxides: from Cystal Chemistry to Physics; Wiley-VCH : **2012**
(2) Maignan, A.; Martin, C.; Pelloquin, D.; Nguyen, N.; Raveau, B. Structural and Magnetic Studies of Ordered Oxygen-Deficient Perovskites LnBaCo2O5+δ, Closely Related to the "112" Structure. *J. Solid State Chem.* **1999**, *142*, 247–260.
(3) Burley, J. C.; Mitchell, J. F.; Short, S.; Miller, D.; Tang, Y. Structural and Magnetic Chemistry of NdBaCo2O5+δ. *J. Solid State Chem.* **2003**, *170*, 339–350.
(4) Vogt, T.; Woodward, P. M.; Karen, P.; Hunter, B. A.; Henning, P.; Moodenbaugh, A. R. Low to High Spin-State Transition Induced by Charge Ordering in Antiferromagnetic YBaCo2O5. *Phys. Rev. Lett.* **2000**, *84*, 2969.
(5) Fauth, F.; Suard, E.; Caignaert, V.; Mirebeau, I.; Keller, L.; Domenges, B. Interplay of Structural, Magnetic and Transport Properties in the Layered Co-based Perovskite LnBaCo2O5 (Ln =Tb, Dy, Ho). *Eur. Phys. J. B* **2001**, *21*, 163.
(6) Barbey, L.; Nguyen, N.; Ducouret, A.; Caignaert, V.; Grenèche, J. M.; Raveau, B. Magnetic Behavior of the "112" Type Substituted Cuprate YBaCoCu1‑xFexO5. *J. Solid State Chem.* **1995**, *115*, 514–520.
(7) Suard, E.; Fauth, F.; Caignaert, V.; Mirebeau, I.; Baldinozzi, G. Charge ordering in the layered Co-based perovskite HoBaCo2O5. *Phys. Rev. B: Condens. Matter Phys.* **2000**, *61*, R11871.
(8) Rautama, E. L.; Boullay, P.; Kundu, A. K.; Caignaert, V.; Pralong, V.; Karpinnen, M.; Raveau, B. Cationic Ordering and Microstructural Effects in the Ferromagnetic Perovskite La0.5Ba0.5CoO3: Impact Upon Magnetotransport Properties. *Chem. Mater.* **2008**, *20*, 2742.
(9) Kundu, A. K.; Rautama, E. L.; Boullay, P.; Caignaert, V.; Pralong, V.; Raveau, B. Spin-Locking Effect in the Nanoscale Ordered Perovskite Cobaltite LaBaCo2O6. *Phys. Rev. B: Condens. Matter Phys.* **2007**, *76*, 184432.
(10) Nakajima, T.; Ichihara, M.; Ueda, Y. New A-site Ordered Perovskite Cobaltite LaBaCo2O6: Synthesis, Structure, Physical Property and Cation Order–Disorder Effect. *J. Phys. Soc. Jpn.* **2005**, *74*, 1572–1577.
(11) Seikh, M. M.; Pralong, V.; Lebedev, O. I.; Caignaert, V.; Raveau, B. The Ordered Double Perovskite PrBaCo2O6: Synthesis, Structure and Magnetism. *J. Appl. Phys.* **2013**, *114*, 013902.
(12) Frontera, C.; Garcia-Munoz, J. L.; Carillo, A. E.; Ritter, C.; Marero, D.; Caneiro, A. Structural and magnetic study of PrBaCo2O5+δ(δ≃0.75) cobaltite. *Phys. Rev. B: Condens. Matter Phys.* **2004**, *70*, 184428.
(13) Streule, S.; Podlesnyak, A.; Sheptyakov, D.; Pomjakushina, E.; Stingaciu, M.; Conder, K.; Medarde, M.; Patrakiev, M. V.; Leodinov, I. A.; Kozhenikov, V. L.; Mesot, J. High-temperature Order-disorder Transition and Polaronic Conductivity in PrBaCo2O5.48. *Phys. Rev. B: Condens. Matter Phys.* **2006**, *73*, 094203.
(14) Frontera, C.; Garcia-Munoz, J. L.; Llobet, A.; Aranda, M. A. Selective Spin-State Switch and Metal-Insulator Transition in GdBaCo2O5.5. *Phys. Rev. B: Condens. Matter Phys.* **2002**, *65*, 180405.
(15) Kusuya, H.; Machida, A.; Moritomo, Y.; Kato, K.; Nishibori, E.; Takata, M.; Sakata, M.; Nakamura, A. Insulator Transition of Tb2Ba2Co4O11. *J. Phys. Soc. Jpn.* **2001**, *70*, 3577–3580.
(16) Plakhty, V. P.; Chernenkov, Y. P.; Barilo, S. N.; Podlesnyak, E.; Pomjakushina, E.; Moskvin, E. V.; Gavrilov, S. V. Spin Structure and Magnetic Phase Transitions in TbBaCo2O5.5. *Phys. Rev. B: Condens. Matter Phys.* **2005**, *71*, 214407.
(17) Jorgensen, J.; Keller, L. Magnetic Ordering in HoBaCo2O5.5. *Phys. Rev. B: Condens. Matter Phys.* **2008**, *77*, 024427.
(18) Moritomo, Y.; Akimito, Y.; Takeo, M.; Machida, A.; Nishibori, E.; Takata, M.; Sakata, M.; Ohoyama, K.; Nakamura, A. Metal-Insulator Transition Induced by a Spin-State Transition in TbBaCo2O5+δ (δ = 0.5). *Phys. Rev. B: Condens. Matter Phys.* **2000**, *61*, R13325.
(19) Khalyavin, D. D.; Argyriou, D. N.; Amann, U.; Yaremchenko, A. A.; Kharton, V. V. Spin-state ordering and magnetic structures in the cobaltites YBaCo2O5+δ (δ = 0.50 and 0.44). *Phys. Rev. B: Condens. Matter Phys.* **2007**, *75*, 134407.
(20) Akahoshi, D.; Ueda, Y. Oxygen Nonstoichiometry, Structures, and Physical Properties of YBaCo2O5+x (0.00 ≤ x ≤ 0.52). *J. Solid State Chem.* **2001**, *156*, 355–363.

(21) Fauth, F.; Suard, E.; Caignaert, V.; Mirebeau, I. Spin-State Ordered Clusters in the Perovskite NdBaCo2O5.47. *Phys. Rev. B:Condens. Matter Phys.* **2002,** *66*, 184421.
(22) Seikh, M. M.; Simon, C.; Caignaert, V.; Pralong, V.; Lepetit, B.; Boudin, S.; Raveau, B. New Magnetic Transitions in the Ordered Oxygen-Deficient Perovskite LnBaCo2O5.50+d. *Chem. Mater.* **2008,** *20*, 231–238.
(23) Taskin, A. A.; Lavrov, A. N.; Ando, Y. Transport and Magnetic Properties of GdBaCo2O5+x Single Crystals: A Cobalt Oxide with Square-Lattice CoO2 Planes Over a Wide Range of Electron and HoleDoping. *Phys. Rev. B: Condens. Matter Phys.* **2005,** *71*, 134414.
(24) Soda, M.; Yasui, Y.; Fujita, T.; Miyashita, T.; Sato, M.; Kakurai, K. Magnetic Structures of High Temperature Phases of TbBaCo2O5.5.*J. Phys. Soc. Jpn.* **2003,** *72*, 1729.
(25) Roy, S.; Dubenko, I. S.; Khan, M.; Condon, E. M.; Craig, J.; Ali, N.; Liu, W.; Mitchell, B. S. Magnetic Properties of Perovskite-Derived Air-synthesized RBaCo2O5+δ (R = La,Ho) compounds. *Phys. Rev. B:Condens. Matter Phys.* **2005,** *71*, 024419.
(26) Karen, P.; Woodward, P. M. Synthesis and Structural Investigations of the Double Perovskites REBaFe2O5+w (RE = Nd, Sm). *J. Mater. Chem.* **1999,** *9*, 789.
(27) Elzubair, A.; El Massalami, M.; Domingues, P. H. On the Structure and Magnetic Properties of the Series RBa2Fe3O8+x (R = La,Nd,Sm,Gd). *Physica B* **1999,** *271*, 284.
(28) Karen, P.; Woodward, P.; Santhosh, P.; Vogt, T.; Stephens, P. W.; Pagola, S. Verwey Transition under Oxygen Loading in RBaFe2O5+w (R = Nd and Sm). *J. Solid State Chem.* **2002,** *167*, 480.
(29) Moritomo, Y.; Hanawa, M.; Ohishi, Y.; Kato, K.; Nakamura, J.; Karppinen, M.; Yamauchi, H. Physical Pressure Effect on the Charge- Ordering Transition of BaSmFe2O5.0. *Phys. Rev. B: Condens. Matter Phys.* **2003,** *68*, 060101.
(30) Woodward, P. M.; Karen, P. Mixed Valence in YBaFe2O5. *Inorg.Chem.* **2003,** *42*, 1121–1129.
(31) Spiel, C.; Blaha, P.; Schwarz, K. Density functional calculations on the charge-ordered and valence-mixed modification of YBaFe2O5. *Phys. Rev. B: Condens. Matter Phys.* **2009,** *79*, 115123.
(32) Lekse, J. W.; Natesakhawat, S. A.; Matranga, C.; Alfonso, D. An Experimental and Computational Investigation of the Oxygen Storage Properties of BaLnFe2O5+δ and BaLnCo2O5+δ (Ln = La, Y)Perovskites. *J. Mater. Chem. A* **2014,** *2*, 2397.
(33) Lebedev, O; I. Turner, S.; Caignaert ,V; Cherepanov, V. A.;Raveau ,B. Exceptional Layered Ordering of Cobalt and Iron in Perovskites. *Chem. Mater.* **2016,** *28*, 2907.
(34) Karen, P.; Kjekshus, A.; Huang, Q.; Lynn, J. W.; Rosov, N.; Sora, I. N.; Karen, V. L.; Mighel, A. D.; Santoro, A. Neutron and X-Ray Powder Diffraction Study of RBa2Fe3O8+wPhases. *J. Solid State Chem.* **1998,** *136*, 21.
(35) Huang, Q.; Karen, P.; Karen, V. L.; Kjekshus, A.; Lynn, J. W.; Mighell, A. D.; Rosov, N.; Santoro, A. Neutron-Powder-Diffraction Study of the Nuclear and Magnetic Structures of YBa2Fe3O8 at RoomTemperature. *Phys. Rev. B: Condens. Matter Phys.* **1992,** *45*, 9611.
(36) Karen, P.; Suard, E.;Fauth, F. Crystal Structure of Stoichiometric YBa2Fe3O8. *Inorg.Chem.* **2005,** *44*, 8170.
(37) Garcia-Gonzalez, E.; Parras, M.; Gonzalez-Calbet, J. M.; Vallet- Regi, M. A New ″123″ Family: LnBa2Fe3Oz: I. LN = Dy, Ho. *J. SolidState Chem.* **1993,** *104*, 232.
(38) Collins, C.; Dyer, M. S.; Demont, A.; Chater, P. A.; Thomas, M. F.; Darling, G. R.; Claridge, J. B.; Rosseinsky, M. J. Computational Prediction and Experimental Confirmation of B-site Doping inYBa2Fe3O8. *Chemical Science* **2014,** 5, 1493.
(39) Volkova, N. E.; Lebedev, O. I.; Gavrilova, Y. L.; Turner, S.; Seikh, M. M.; Gauquelin, N.; Caignaert, V.; Cherepanov, V. A.; Raveau, B.; Van Tendeloo, G. Nanoscale Ordered Oxygen DeficientQuintuple Perovskite Sm2-εBa3+εFe5O15-δ. *Chem. Mater.* **2014**, 26, 6303.
(40) Kundu, A. K.; Lebedev, O. I.;Volkova, N. E.; Caignaert, V.; Seikh, M. M.; Cherepanov, V. A.; Raveau, B. Quintuple Perovskites Ln2Ba3Fe5-xCoxO15-δ: Nanoscale Ordering and Unconventional Magnetism. *J. Mater. Chem. C* **2015**, *3*, 5398.
(41) Kundu, A. K.; Mychinko, M. Y.;Caignaert, V.; Lebedev, O. I.; Volkova, N. E.; Deryabina, K. M.; Cherepanov, V.; Raveau, B. Quintuple Perovskite Phasoids in the system Nd2‐εBa3+ε(Fe,Co)5O15−δ. *J. Solid State Chem.* **2015**, *231*, 36–41.
(42) Solanki, V.; Das, S.; Kumar, S.; Seikh, M. M.; Raveau, B.; Kundu, A. K. Crucial role of sol-gel synthesis in the structural and magnetic properties of LaFe0.5(Co/Ni)0.5O3 perovskites. *J. Sol-Gel Sci. Technol.* **2017**, *82*, 536-540.
(43) Lee, J. H.; Jeong, Y. K.; Park, J. H.; Oak, M.; Jang, H. M.; Son, J. Y.; Scott, J. F. Spin-Canting-Induced Improper Ferroelectricity and Spontaneous Magnetization Reversal in SmFeO3. *Phys. Rev. Lett.* **2011**, *107*, 117201.
(44) Johnson, R. D.; Terada,N.;Radaelli, P. G.Comment on "Spin-Canting-Induced Improper Ferroelectricity and Spontaneous Magnetization Reversal in SmFeO3".*Phys. Rev. Lett.* **2012**, 108, 219701.
(45) Lee, J. H.; Jeong, Y. K.; Park, J. H.; Oak, M.; Jang, H. M.; Son, J. Y.; Scott, J. F. Lee et al Reply.*Phys. Rev. Lett.***2012**, *108*, 219702.
(46) Kuo, C.-Y.; Drees, Y.; Fernández-Díaz, M. T.; Zhao, L.; Vasylechko, L.; Sheptyakov, D.; Bell, A. M. T.; Pi, T. W.; Lin, J.-H.; Wu, M-K.; Pellegrin, E.; Valvidares, S. M.; Li, Z. W.; Adler, P.; Todorova, A.; Küchler, R.; Steppke, A.; Tjeng, L. H.; Hu, Z.; Komarek, A. C. k=0Magnetic Structure and Absence of Ferroelectricity in SmFeO3. *Phys. Rev. Lett.* **2014**, *113*, 217203.
(47) Ivanovaa, N. B.; Kazaka, N. V.; Michelb, C. R.; Balaeva, A. D.; Ovchinnikova, S. G. Low-Temperature Magnetic Behavior of the Rare-Earth Cobaltites GdCoO3 and SmCoO3.*Physics of the Solid State.* **2007,** *49*, 2126–2131.
(48) Tachibana, M.; Yoshida, T.;Kawaji, H.;Atake, T.; Muromachi, E. T. Evolution of electronic states in RCoO3(R=rareearth): Heat capacity measurements.*Phys. Rev. B* **2008**, *77*, 094402.
(49) Tan, H. ; Verbeek, J. ; Abakumov, A. ; Tendeloo, G. V. Oxidation state and chemical shift investigation in transition metal oxides by EELS. *Ultramicroscopy* **2012**, *116*, 24-33.
(50) Colliex, C.;Manoubi, T.; Ortiz, C.Electron-energy-loss-spectroscopy near-edge fine structures in the iron-oxygen system.*Phys.Rev.B*, **1991,** 44, 11402.
(51) Wang, Z. L.; Yin, J.S.; Jiang, Y.D. EELS analysis of cation valence states and oxygen vacancies in magnetic oxides. *Micron* **2000**, *31*, 571–580.

(52) Yuan, S.J.; Ren, W.; Hong, F.; Wang, Y.B.; Zhang, J.C.; Bellaiche, L.; Cao, S.X.; Cao, G. Spin switching and magnetization reversal in single-crystal NdFeO3. *Phys. Rev. B* **2013,** *87*, 184405.

(53) Anil Kumar, P. S.; Santhosh Nair, P. N.; Joy, P. A.; Date, S. K. Evolution of the cluster glass system $La_{0.5}Sr_{0.5}CoO_3$. *J. Mater. Chem.* **1998,** *8*, 2245.